\documentclass{achemso}
\setkeys{acs}{articletitle = true}

\usepackage{graphicx}
\usepackage{amsmath,amssymb,latexsym,xspace}

\usepackage{dcolumn}
\usepackage{bm}
\usepackage{subfigure}
\usepackage{epsfig}
\usepackage{epstopdf}
\usepackage{mathptmx}

\SectionNumbersOn

\author{Kevin Conley}
\affiliation[Aalto University]
{Department of Applied Physics, COMP Centre of Excellence, Aalto University School of Science, P.O. Box 11100, FI-00076 Aalto, Finland}
\alsoaffiliation[Aalto University2]{Department of Applied Physics, QTF Centre of Excellence, Aalto University School of Science, P.O. Box 11100, FI-00076 Aalto, Finland}
\author{Neha Nayyar}
\affiliation[UCF]
{Department of Physics, University of Central Florida, Orlando, Florida, USA}
\author{Tuomas P. Rossi}
\affiliation[Aalto University]
{Department of Applied Physics, COMP Centre of Excellence, Aalto University School of Science, P.O. Box 11100, FI-00076 Aalto, Finland}
\alsoaffiliation[Chalmers]{Department of Physics, Chalmers University of Technology, Gothenburg, Sweden}
\author{Mikael Kuisma}
\affiliation[UoJ]
{Department of Chemistry, Nanoscience Center, University of Jyv\"{a}skyl\"{a}, FI-40014 Jyv\"{a}skyl\"{a}, Finland}
\author{Volodymyr Turkowski}
\affiliation[UCF]
{Department of Physics, University of Central Florida, Orlando, Florida, USA}
\author{Martti Puska}
\affiliation[Aalto University]
{Department of Applied Physics, COMP Centre of Excellence, Aalto University School of Science, P.O. Box 11100, FI-00076 Aalto, Finland}
\author{Talat S. Rahman}
\affiliation[UCF]
{Department of Physics, University of Central Florida, Orlando, Florida, USA}
\alsoaffiliation[Aalto University]
{Department of Applied Physics, COMP Centre of Excellence, Aalto University School of Science, P.O. Box 11100, FI-00076 Aalto, Finland}

\email{talat.rahman@ucf.edu}

\title{Plasmon excitations in mixed metallic nanoarrays}

\begin{document}

\begin{abstract}
We study the plasmonic properties of arrays of atomic chains which comprise noble (Cu, Ag, and Au) and transition (Pd, Pt) metal atoms using time-dependent density-functional theory. We show that the response to the electromagnetic radiation is related to both physics, the geometry-dependent confinement of sp-valence electrons, and chemistry, the energy position of d-electrons in the different atomic species and the hybridization between d and sp electrons. As a result it is possible to tune the position of the surface plasmon resonance, split it to several peaks, and eventually achieve broadband absorption of radiation. Mixing the arrays with transition metals can strongly attenuate the plasmonic behaviour. We analyze the origin of these  phenomena and show that they arise from rich interactions between single-particle electron-hole and collective electron excitations. The tunability of the plasmonic response of arrays of atomic chains, which can be realized on solid surfaces, opens wide possibilities for their applications. In the present study we obtain guidelines how the desired properties can be achieved.

\end{abstract}

\section{Introduction}\label{intro}
The optical properties of nanosystems are highly sensitive to their size, shape, and the chemical composition, and can dramatically differ from those of their bulk cousins. Over the past decade, a wide variety of plasmonic structures based on gold and silver have been fabricated to manipulate the light absorption at the nanometer scale for novel applications and basic research of physical phenomena.~\cite{quinten1998electromagnetic,solis2012electromagnetic,cortie2011synthesis,rycenga2011controlling} These applications rely to some extent on the ability to tune the nanoparticle plasmon resonances, which has played a crucial role in stimulating the current interest in nanoplasmonics. In order to be able to intelligently tune the system properties, including the absorption spectrum, it is essential to understand its dependencies on the physical and chemical parameters. 

An early theoretical study on the collective excitations in a few-atom sodium clusters was carried out using Time-Dependent Density Functional Theory (TD-DFT) by Kummel et al.~\cite{kummel2001collectivity} who showed that collective excitations exist even for very small clusters. More recent work addresses the mechanism of the collective excitations and field enhancement in higher dimension clusters using TD-DFT.~\cite{song2011quantum,yasuike2011collectivity,yin2012quantum} Ma et al. studied the sensitivity of plasmon resonance in Au nanoparticles and their dimers as a function of the particle size and the inter-particle distance.~\cite{ma2012plasmon} Studies of the plasmon excitations in two-dimensional planar Na structures~\cite{yin2012collectivity,wang2012plasmon} or MoS$_2$ nanoflakes~\cite{rossi2017effect} reveal the importance of dimensionality in the formation and development of the plasmon peaks. 

Scanning Tunneling Microscope (STM) experiments by Nilius et al.~\cite{nilius2002development} demonstrated the development of 1D band structure in Au chains on NiAl(110) when the number of atoms in the chain exceeds 10. Inspired by this experimental finding, theoretical calculations have also predicted the presence of collective plasmon modes in a few-atom chains of several metallic elements: Na,~\cite{yan2007end} Ag,~\cite{yan2008plasmon} and Au~\cite{lian2009density} (for over-review, see, e.g., Ref.~\cite{guidez2014quantum}). Experimental observation of such collective excitations requires the chains to be grown on a substrate that does not quench them rapidly. While there are theoretical indications that the NiAl(110) surface does not affect the electronic properties of the Au chains,~\cite{calzolari2005first} its metallic nature precludes a short lifetime for any plasmon excitation. On the other hand, it is possible to grow Au chains and wires on semiconductor substrates such as Si(557)~\cite{barke2009coverage}, Ge(001)~\cite{wang2004scanning} and quartz~\cite{wei2004plasmon} which may be amenable for capturing the plasmon effects (especially if the substrate bandgap is much larger than the chain-plasmon and other excitation energies of interest). In a previous study of some of the present authors~\cite{nayyar2012optical} the role of transition metal (TM) doping in the generation of plasmon modes in single Au chains was examined. It was found that doping leads to several new excitations in the absorption spectrum reflecting changes in the potential around the TM atom and  collective effects of the ``localized'' (TM d-) and delocalized (s) electrons. Indeed, the mutual effects of both electronic subsystems may be nontrivial, in particular leading to a change of the spectral function of the localized electrons~\cite{freericks2005f} or to local electronic resonances around the dopant atom, as was shown experimentally in the case of Pd-doped Au chains on NiAl(110).~\cite{nilius2005tailoring} This complexity opens the door to new opportunities for tuning the optical properties by combining noble and transition metal nanostructures. 

Besides changing the composition of nanostructures, the optical spectra can be tuned by changing the size, shape, and geometry. In the present study, we consider an arrangement of small atomic chains and arrays. The optical properties of both pure and ``mixed'' arrays of homonuclear chains are calculated with TD-DFT. In order to understand optical properties, it is essential to analyze the nature of different features in the excitation spectra. This analysis is also the first step in systematic tuning of nanostructures in order to achieve desired properties such as strong resonance peaks or spreading of the absorption intensity over certain wavelength regions. In the analysis we use Transition Contribution Maps (TCMs) to identify how the individual Kohn-Sham (KS) transitions collectively contribute to the given photoabsorption peaks.~\cite{he2010first,malola2013birth} The excitation spectra and TCMs can be constructed from both Casida calculations~\cite{casida1995recent} and the better scaling time-propagation TD-DFT.~\cite{rossi2017kohn} Our results and analyses show a strong interplay between the physical (geometry-dependent electron confinement) and chemical (element-dependent electronic structure) effects parallel to that between the single-particle and collective excitations. For example, we show that in mixed arrays TM atom chains may quench the plasmon mode dwelling in noble metal chains. 

The organization of the paper is as follows. In Section~\ref{methods}, we describe the detailed geometries of the nanoarrays and our computational methods employed in TD-DFT calculations and in their analysis. In Section~\ref{results:pure}, we analyze the plasmons formed in pure atomic arrays. While our results are consistent with previously published atomic systems~\cite{wang2012plasmon,piccini2013gold,guidez2014quantum}, we provide a detailed understanding via analysis with TCM. The TCMs visualize the quantum effects within the array, such as the dependence of the plasmon frequency on array size. Additionally the pure systems are the foundation on which to extend these methods to more challenging mixed systems which require quantum mechanical analysis. In Section~\ref{results:mixed}, we show that the plasmon in mixed arrays can either be maintained or destructively quenched by the second chain. 

\section{Systems studied and methods for their modelling} \label{methods}
First, atomic wires comprising noble (Au, Ag, Cu) or transition (Pd, Pt, Ni, Fe, Rh) metal atoms with length varying from 2 to 19 atoms are constructed. Then, arrays consisting of up to eight chains are assembled to form pure or mixed arrays of homonuclear chains (see Figure~\ref{fig:arraysize_spectra}b for a homonuclear array). To mimic the supported nanostructures, a square (simple cubic) lattice is assumed for planar (rod-like) arrays and the bond length is set to 2.89 \AA, as measured in experimentally realized Au chains on the NiAl(110) substrate~\cite{nilius2002development}. The obtained spectra were insensitive to changes in the bond length within experimental error. 

The electronic structures of the nanostructures are calculated within DFT using the solid-state modified GLLB-SC exchange-correlation potential~\cite{kuisma2010kohn} which accurately describes the energy level positions of the d-states in noble metals~\cite{Yan2011First, Yan2012}. We employ the GPAW code package~\cite{gpaw,Enkovaara2010,gpawTDDFT,Kuisma2015} based on the projector-augmented wave (PAW) method~\cite{blochl1994pe} and utilizing the ASE package~\cite{ase-paper}. Electronic wavefunctions are expanded as linear combinations of atomic orbitals (LCAO)~\cite{larsen2009localized} with less than 0.05 eV Fermi-Dirac smearing of the occupancy number and the electron density and potentials are represented on a real space grid with a grid spacing of 0.3 \AA. The molecules are surrounded by at least 6 \AA{} of vacuum and the Hartree potential is evaluated on a larger and coarser grid with at least 48 \AA{} of vacuum. 

The optical response of the atomic wires and arrays is calculated using the LCAO time propagation (TP) TD-DFT code~\cite{Kuisma2015} with weak $\delta$-pulse perturbation \cite{Yabana1996}. In this approach, the electron wave functions are evolved after an external electric (in dipole approximation) $\delta$-pulse along the long wire axis. The time-dependent induced density provides the time-dependent dipole moment from which the dynamical polarizability and the ensuing frequency-dependent photoabsorption spectrum are determined. The photoabsorption spectrum with Gaussian broadening of $\sigma$ = 0.07 eV is sufficiently obtained using a time step of 10 as for a total propagation time of 30 fs. 

The induced density matrix between occupied and unoccupied KS states and the corresponding dipole matrix elements allows the photoabsorption spectrum to be decomposed into contributions from individual discrete electron-hole transitions with well-defined relative weights.~\cite{rossi2017kohn} The decomposition is represented as a two-dimensional Transition Composition Map (TCM), such as the plasmonic absorption excitation of a 14~x~2 atom Au array at $\omega$ = 1.31 eV shown in Figure~\ref{fig:2xpureTCM}a. The absorption energy corresponds to the ascending probe line, $\omega = \epsilon_u -\epsilon_o$, where $\epsilon_u$ and $\epsilon_o$ are energies within the unoccupied and occupied regions of the single-particle energies and given on the horizontal and vertical axis of TCM, respectively. The different discrete electron-hole transitions are denoted by red and blue spots on the TCM plane such that red (blue) spots correspond to the transitions which have a positive (negative) contribution to the photoabsorption. The intensity of the spot at the location of every electron-hole transition is proportional to the relative weight of the electron-hole transition, whereas the extent is given by the two-dimensional Gaussian broadening function ($\sigma_\mathrm{TCM}$ = 0.04 eV). For convenience, the density of states (DOS) is shown along the top and side, and colored to indicate the sp and d characters of the KS states. 

The great value of TCMs is that they visualize clearly whether there are low-energy electron-hole transitions which collectively form a plasmonic excitation at $\omega$ (red spots below the $\omega$ probe line).~\cite{Yannouleas1992, bernadotte2013plasmons, rossi2017kohn}  The excitation energy comprises the energies of individual transitions and the electron-electron interaction contribution described by TD-DFT.~\cite{zhang2017identify} Negative contributions may cause damping of the plasmonic excitation (blue spots, typically above the line $\omega = \epsilon_u -\epsilon_o$). Moreover, a change from negative (blue) to positive (red) in the contribution of a transition close to the probe line can result in plasmon splitting or fragmentation as will be demonstrated for mixed arrays.~\cite{Yannouleas1989, rossi2017kohn} Similar to the photoabsorption decomposition into different electron-hole contributions, the induced electron density can be decomposed into partial densities corresponding to the different electron-hole transition contributions. These real-space representations show their value in the following discussion. More theoretical and and practical details about these analysis tools can be found in a recent article~\cite{rossi2017kohn}.

Some of the nanowire systems were also calculated using Gaussian 03~\cite{g03} with a B3PW91 hybrid functional~\cite{becke1993density,perdew1986accurate,perdew1991electronic} and a LanL2DZ basis set~\cite{hay1985ab}. The main features of the absorption spectra were consistent with the LCAO-TP-TD-DFT calculations~\cite{nehathesis}. All the results presented and analyzed in this paper are obtained with the LCAO-TP-TD-DFT method. 

\section{Results and Discussion} \label{results}
In this Section, the optical responses of pure and mixed arrays are considered. The effects of the finite size and s(p)-d hybridization of electronic states between similar and different types of atoms on the valence electron structure and thereby on the plasmonic response will be discussed. 

To date, there has been a lively discussion on whether these strong absorption modes in small nanosystems or molecules are plasmons or whether they are single-particle excitations.~\cite{Puska1985, Yannouleas1991, yasuike2011collectivity, piccini2013gold, bernadotte2013plasmons, de2015plasmonic,krauter2015identification,zhang2017identify} For example, Piccini et al.~\citep{piccini2013gold} concluded based on the fact that only one KS transition contributes to the strong absorption peak of atomic Au chains that the excitation is single-particle-like in nature. On the other hand, Bernadotte et al.~\cite{bernadotte2013plasmons} showed that some of the excitations in nanostructures are collective by studying the wave vector dependence of excitations and scaling the electron-electron Coulomb interaction. Although many excitations have single-particle nature, it was concluded that the strongest absorption peaks in atomic Au chains around 1 to 2 eV belong to the collective, plasmonic excitations. After all, a pure single electron excitation depending on Coulomb strength would be a only an indication of self-interaction error. Such aspects can be visualized with TCMs for differentiating collective excitations from single-particle excitations in molecular systems, such as the atomic nanoarrays considered in this paper. Although features of plasmons from macroscopic materials are known to emerge in molecular systems,~\cite{de2015plasmonic} our systems are far from the size of nanoparticles showing clear surface plasmon resonance with induced densities restricted on the particle surface.~\cite{zhang2017identify} The intense absorption modes from collective excitations in molecular systems could be described as molecular plasmons to distinguish them from conventional plasmons. However, for short we use the term plasmon interchangeably.

\subsection{Molecular plasmons in pure arrays}\label{results:pure}
\begin{figure}
\includegraphics[width=0.45\textwidth]{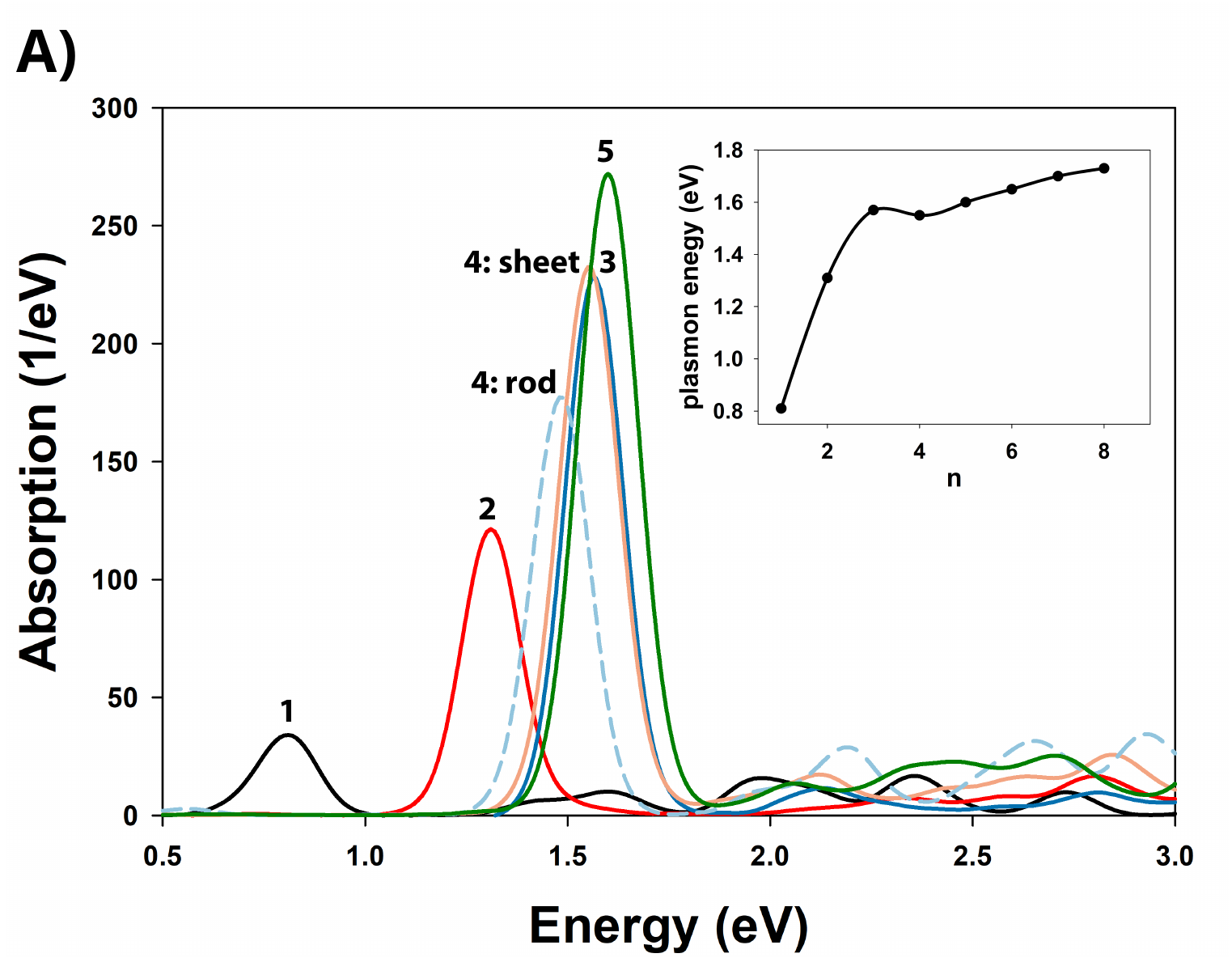}
\includegraphics[width=0.45\textwidth]{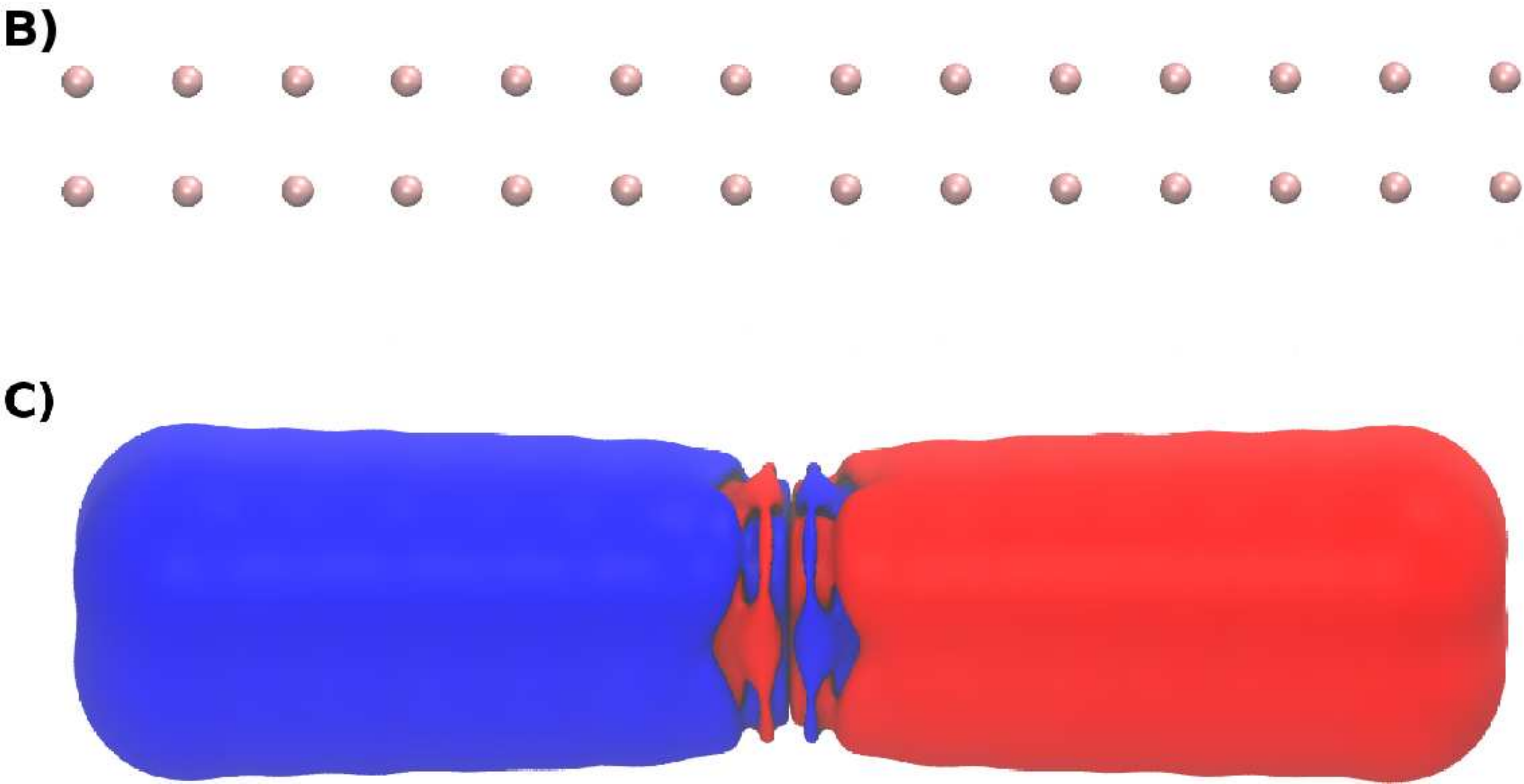}\\
\caption{a) Absorption spectra of Au arrays formed by one to five chains of 14 Au atoms. The arrays of 2, 3, and 5 chains are planar. The 4 chain system is arranged into a 4~x~14 planar array and a 2~x~2~x~14 nanowire. Inset - plasmon energy vs number of chains in a n~x~14 array. b) Atomic structure of the planar 2~x~14 array and c) the total plasmonic transition density for the 2~x~2~x~14 nanowire at $\omega$ = 1.48 eV. }
\label{fig:arraysize_spectra}
\end{figure}

Previous studies of single Au chains revealed that increasing the chain length above approximately 8 to 10 atoms generates a strong absorption mode~\cite{lian2009density,nayyar2012optical,piccini2013gold}. Increasing the number of atoms further intensifies the modes and causes a slowly saturating redshift. In this work, we studied similar trends in Au nanoarrays. The absorption spectra for the single chain of 14 Au atoms as well as for arrays of $n$ chains of 14 Au atoms, where $n$ = 2~-~5, are shown in Figure~\ref{fig:arraysize_spectra}a. All of the arrays are planar ($n$~x~1~x~14) with the exception of the rod-like 2~x~2~x~14 nanowire. As examples of the atomic structures and induced plasmonic densities, we show the 3~x~14 and 2~x~2~x~14 arrays, respectively, in Figure~\ref{fig:arraysize_spectra}b. The plasmon blueshifts toward the visible range (1.7 eV - 3.0 eV) as the number of chains increases in single, double, and triple chain arrays. The blueshift is in accordance with TD-DFT results for Au nanowires for which the plasmon energy increases with the nanowire diameter.~\cite{piccini2013gold} The jump in the plasmon energy is largest from one to two Au chains. This is due to the increase of the valence electron confinement and the average density between the atom chains as seen in Figure~\ref{fig:arraysize_spectra}a inset; classically the plasma frequency increases with the increasing electron density. 

\begin{figure}
\includegraphics[width=0.75\textwidth]{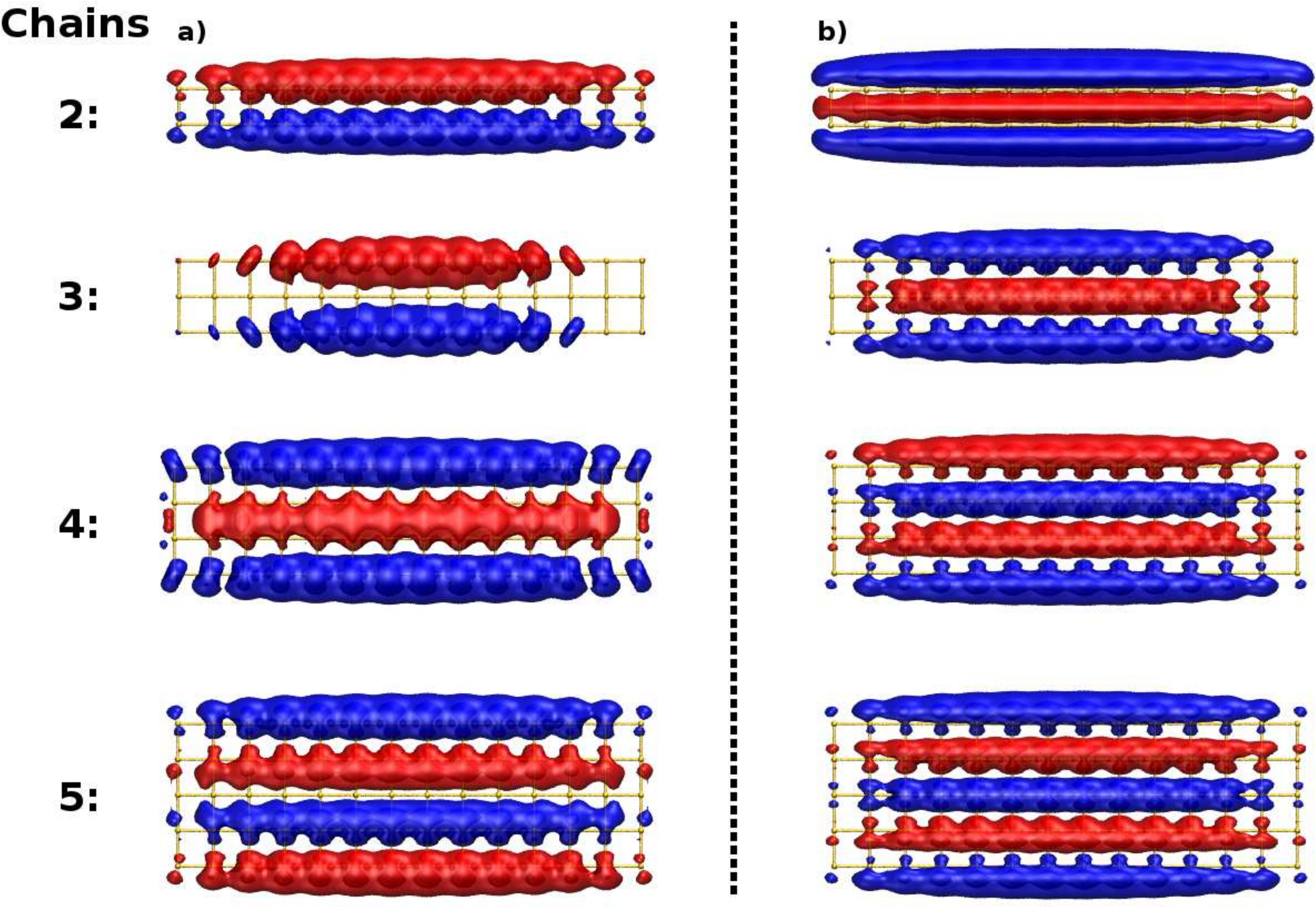}
\caption{\textbf{Schematic of the subband structure in Au arrays.} KS orbitals in Au arrays with different number of chains. a) Highest order subband in occupied KS orbitals. b) Lowest order subband opened in the unoccupied KS orbitals.}
\label{fig:subbands_all}
\end{figure}

Slightly surprisingly the plasmon energy does not increase but slightly decreases between the planar arrays of three to four Au chains. Thereafter, the plasmon energy increases again, although rather moderately and monotonically, up to the array of 8 Au chains. The non-monotonic behavior for the smallest arrays, which does not arise classically, is due to the quantum mechanical confinement of the average electron density between the Au chains and the evolution of the nodal structure of occupied state valence electron wavefunctions perpendicular to the Au chains. ``Subbands'' corresponding to a certain number transverse nodes (with nodal surfaces parallel to the chain axis) are opened and filled with electrons. The details of the finite-size effect have been checked by visually inspecting the KS wavefunctions in different Au arrays. Examples of wavefunctions and a schematic of the nodal structures in different arrays are given in Figures~\ref{fig:subbands_all} and S1, respectively.

Subbands of at most one transverse node are occupied in arrays of two and three Au chains. Although the existence of a nodal plane is expected to result in increased electron spillout and decrease the plasmonic frequency this is not the case here. The electron spillout from a single chain is much larger than that from the double chain. Moreover in the case of the triple chain the nodal plane is far from the edges of the array, i.e., on top of the middle Au chain which results in a compact electronic structure and further large increase in the plasmonic frequency. Arrays of four Au chains are the smallest array which have an open subband of two nodal planes below the Fermi level. These new nodal planes lie between the chains and close to the array edges where they cause the electron density to spillout from between the chains. The effect of the ensuing spillout is strong enough to decrease the plasmonic frequency despite having more electrons. This plasmonic redshift is, in part, related to the unique nodal structure of the three chain array. Unlike the other sizes, the single node in the three chain array is far from the array edges resulting in weak spillout. The weak electron spillout is expressed in the small work function of the triple chain array relative to the other sizes, and shown in the density of states with respect to the vacuum level in Figure S2. Opening more subbands with edge nodes in larger arrays does not have as pronounced of an influence and the change to the work function is minor. Thus the plasmonic frequency increases modestly and monotonically up to arrays of eight chains. 

In a similar manner, the electron spillout from the 2~x~2~x~14 nanowire is larger than that from the planar 4~x~14 array and the plasmonic frequency redshifts from the array to the nanowire. The transverse nodes in the nanowire are between two Au on each face so that the plasmonic frequency is even lower than that of the 3~x~14 array. However, the electron density corresponding to the lowest subband has a tendency to confine in the middle of four Au chains resulting in a plasmonic frequency larger that that for the 2~x~14 array.

\begin{figure}
\includegraphics[width=0.45\textwidth]{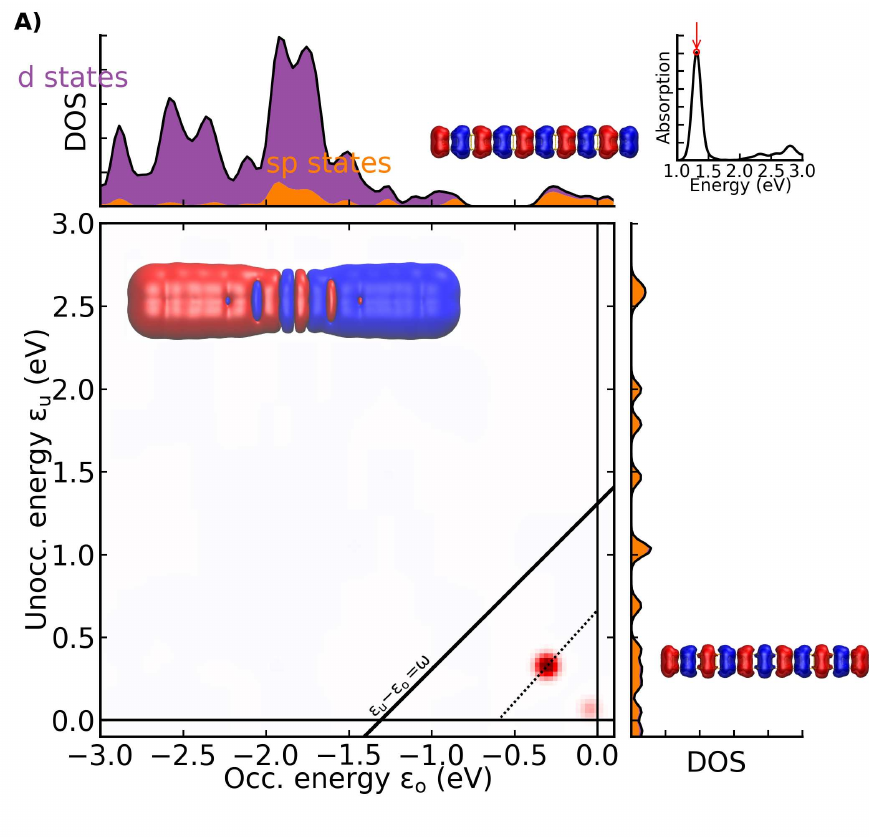}
\includegraphics[width=0.45\textwidth]{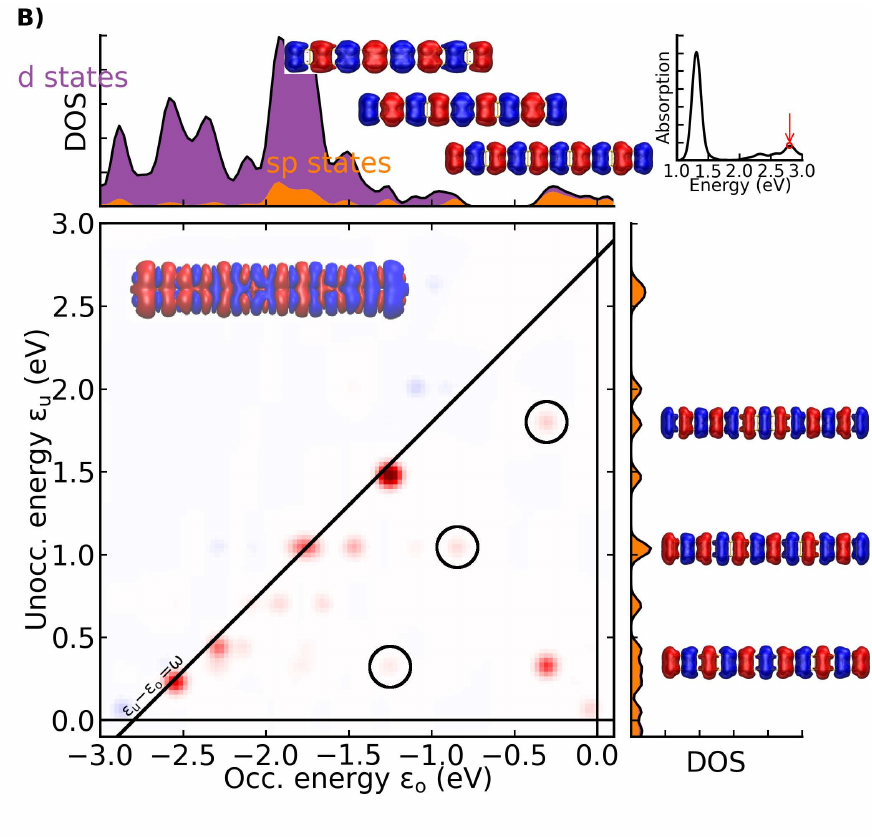}
\caption{Transition Contribution Maps for the 2~x~14 Au array at a) the plasmonic peak at $\omega$ = 1.31 eV and b) at a higher energy peak with contribution from single-particle and collective electron transitions at $\omega$ = 2.80 eV. The KS orbitals involved in the plasmonic transition are shown next to the DOS and the total transition densities are shown in the insets.}
\label{fig:2xpureTCM}
\end{figure}

A more detailed quantum mechanical understanding of the plasmonic absorption of Au arrays in Figure~\ref{fig:arraysize_spectra}a is gained by inspecting the KS electron-hole transition contributions visualized as TCMs. We pay particular focus to the 2~x~14 Au double chain because its characteristics will be compared with those of the simplest and most fundamental mixed arrays comprising only two chains of different atomic species. The TCM of the 2~x~14 Au double chain array for the molecular plasmon at 1.48 eV is given in Figure~\ref{fig:2xpureTCM}a. It has two regions of red illumination corresponding to the two individual electron-hole transitions which form the plasmon. The brighter, i.e. stronger, contribution corresponds to the KS transition between delocalized sp-states with 9 and 10 longitudinal nodes and no transverse nodes. The wavefunctions of these states are shown next to the occupied and unoccupied DOSs, respectively. The difference in the numbers of longitudinal nodes is one, which produces a dipolar total induced density, shown in the inset. The HOMO $\rightarrow$ LUMO transition corresponds to the weakly illuminated red spot in the lower right corner of the TCM. In addition to the longitudinal nodes, the HOMO and LUMO are part of the subband containing one transverse node. 

The TCM of the 2~x~14 Au double chain at the molecular plasmon energy is remarkably different from that of the single chain of 14 Au atoms shown in Figure S3. In accordance with previous predictions for a single Au chain,~\cite{piccini2013gold} the HOMO $\rightarrow$ LUMO transition contributes most strongly to the molecular plasmon. It is evident from the DOSs that the introduction of the second chain pushes the d electron levels downwards relative to the Fermi level. In the double chain, the sp-states are split into bonding and antibonding states between the chains so that the states closest to the Fermi level are antibonding. This hybridization produces the subband structure in the double chain arrays discussed above. The occupied subband with highest order and zero longitudinal nodes is shown in Figure~\ref{fig:subbands_all}a. The strongest induced dipole moment is, however, between bonding states and their KS energies differ by about 0.6 eV. 

Both the TCM of the strong absorption peak in the double chain array in Figure~\ref{fig:2xpureTCM}a and the TCM for the single chain in Figure S3 show one dominant electron-hole transition contribution. The plasmon resonance energy, $\omega$, is significantly larger than the KS eigenvalue difference of the corresponding transition. This shift is a result of the strong influence of the electron-electron interaction taken into account in TD-DFT, and obtained as a gradual evolution in the scaling method.~\citep{bernadotte2013plasmons,krauter2015identification} In this respect, these resonances can be attributed as molecular plasmon excitations despite being comprised of only one dominant transition.

\begin{figure}
\includegraphics[width=0.32\textwidth]{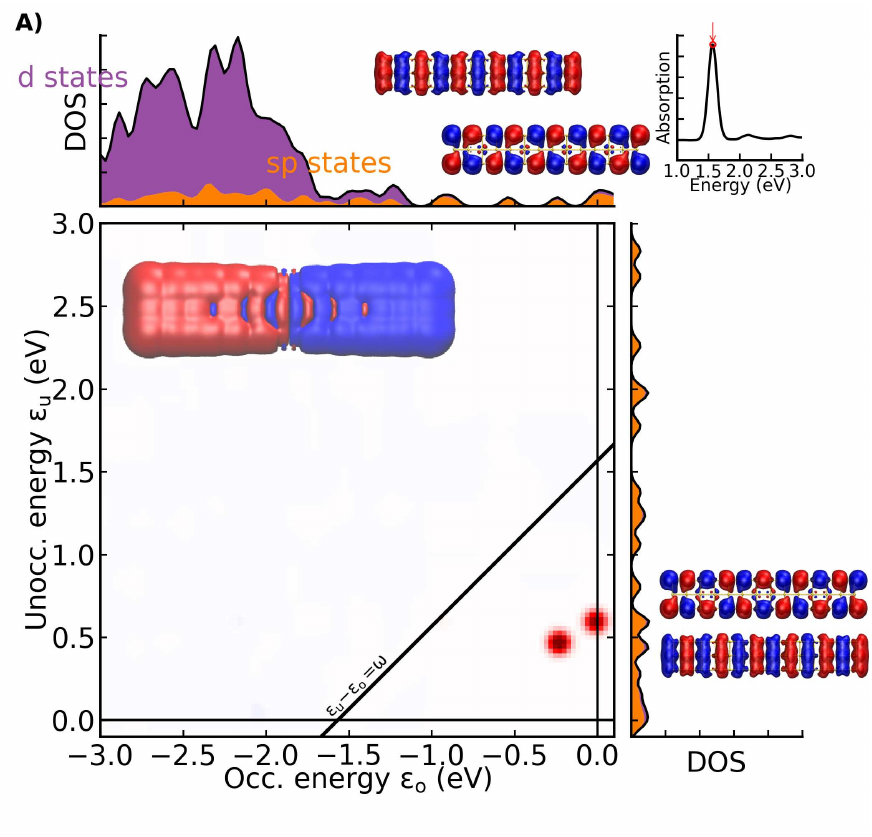}
\includegraphics[width=0.32\textwidth]{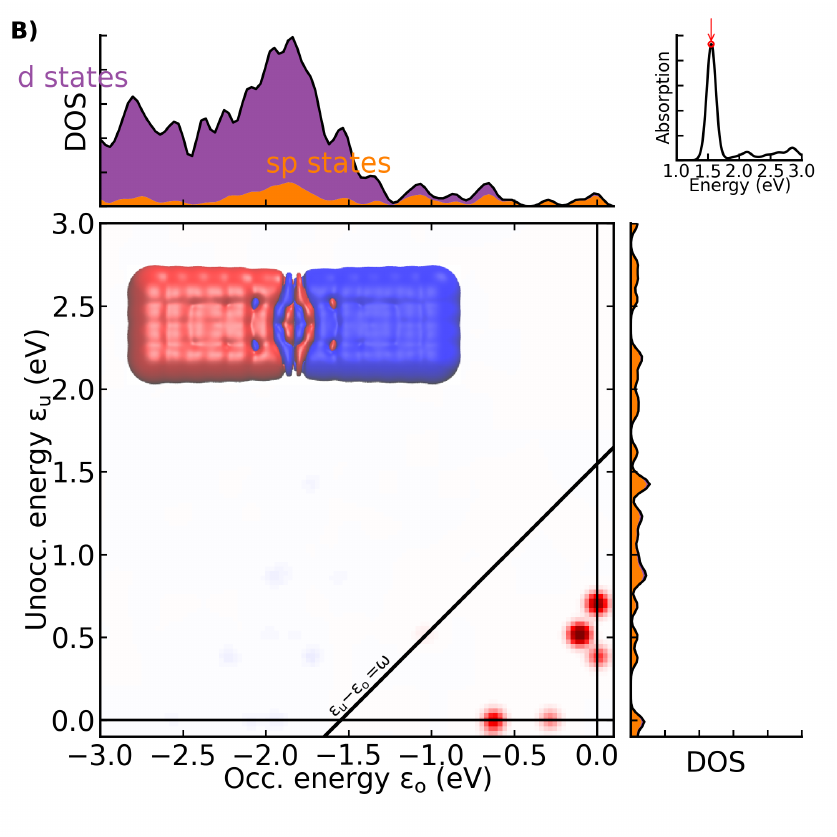}
\includegraphics[width=0.32\textwidth]{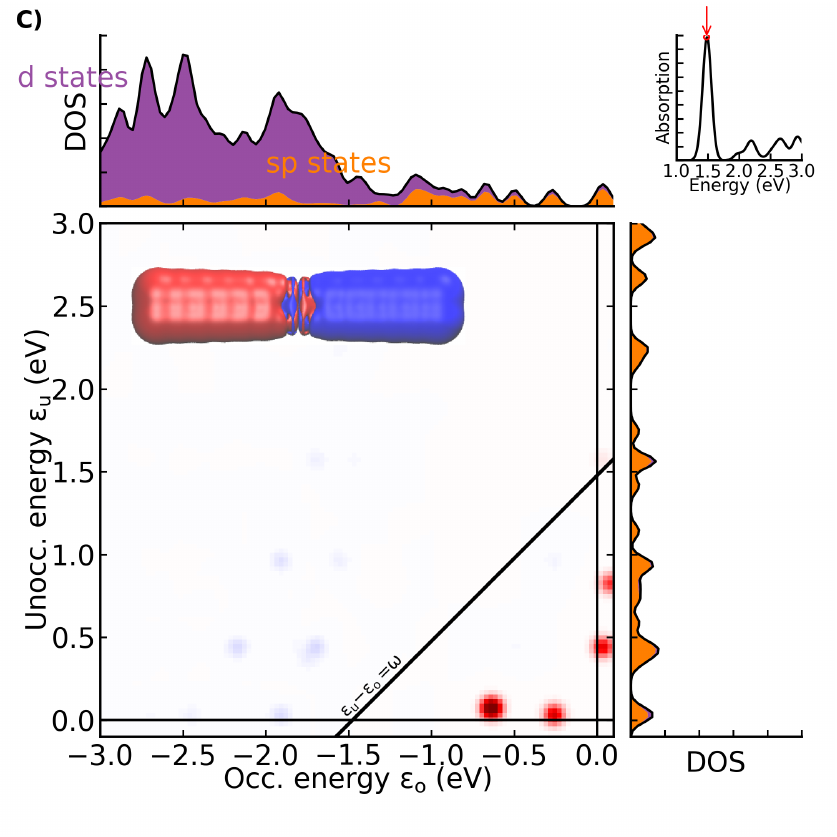}
\caption{Transition Contribution Maps a) for the 3~x~14 Au array at the plasmonic peak at $\omega$ = 1.57 eV, b) for the planar 4~x~14 Au array at the plasmonic peak at $\omega$ = 1.55 eV, and c) for the rod-like 2~x~2~x~14 Au nonowire at the plasmonic peak at $\omega$ = 1.48 eV. Total transition densities are shown in the insets.}
\label{fig:larger_pureTCM}
\end{figure}

The TCMs for the molecular plasmons in wider Au arrays are shown in Figure~\ref{fig:larger_pureTCM}. The transverse nodes and subband structure of the larger arrays permit more sp-type KS states close to the Fermi level. Filling the subbands increases the number of electron-hole transitions contributing to the molecular plasmons of larger arrays. There are two transitions in the TCM of the 2~x~14 and 3~x~14 arrays; each filled subband provides one transition and produces a strong induced dipole density. These KS orbitals are provided next to the respective TCMs. Similarly, in larger arrays, transitions occur from each of the filled subbands (in addition to transitions between fractionally occupied states). In this manner, the collective nature of the molecular plasmon (in this instance collectivity refers to the number of contributing electron-hole transitions) increases with the number of chains in Au arrays.

We also notice that the d levels of the triple chain 3~x~14 Au array, as seen in Figure~\ref{fig:2xpureTCM}a, are lower in energy than those of the double chain array. This stability reflects the more compact electronic structure of the former system discussed earlier. In both the sheet-like and rod-like arrays of four Au chains the stronger splitting between bonding and antibonding d states pushes the d levels upwards. However, they are still lower in energy than in the case of the single Au chain (See Figure S3) which minimizes the plasmon damping.

Faint blue spots appear above the $\omega$ probe line in the TCM of the plasmonic excitation in Figure~\ref{fig:larger_pureTCM}(b,c). Weak KS transitions into the d-states screen the plasmonic excitation and strengthen as the plasmon blueshifts with increased number of chains in the array. This screening, although weak, slows the increase of the plasmon intensity despite the increased collective effects in Figure~\ref{fig:arraysize_spectra}. Previously it was shown that the plasmon intensity of Na arrays, which do not have the d-band, increases with array size up to 16~x~16.~\cite{wang2012plasmon}

In the above TCMs for molecular plasmonic excitations, the main contributing KS transitions appear as red spots below the probe line corresponding to the plasmon resonance energy, $\omega = \epsilon_u -\epsilon_o$. These transitions remain weakly illuminated when probing the higher energy resonances. We present the TCM of the double chain Au array at an excitation peak of 2.80 eV as an example in Figure~\ref{fig:2xpureTCM}b. Higher harmonic molecular plasmons also contribute to this absorption peak and are circled. These electron-hole transitions occur between states where the difference in the number of longitudinal nodes is three. Similar higher order molecular plasmons have been identified in atomic chains by Bernadotte et al. using the $\lambda$-scaling method.~\cite{bernadotte2013plasmons} The TCM clearly shows that the photoabsorption peak has contributions from both plasmons below the probe line and single-particle excitations on the probe line. The induced density in the inset of the figure is a product of the mixing and constructive interference between the dipole moments of each contribution. This mixing of plasmonic and single-particle transitions offers a possibility of generating hot electrons at the dense d-states. There are rather few states and possible transitions between them, and so hot electron generation is not expected to be an intense process in double chains. Arrays of several atomic chains would be more effective.

\subsection{Molecular plasmons in mixed arrays}\label{results:mixed}
\begin{figure}
\includegraphics[width=\textwidth]{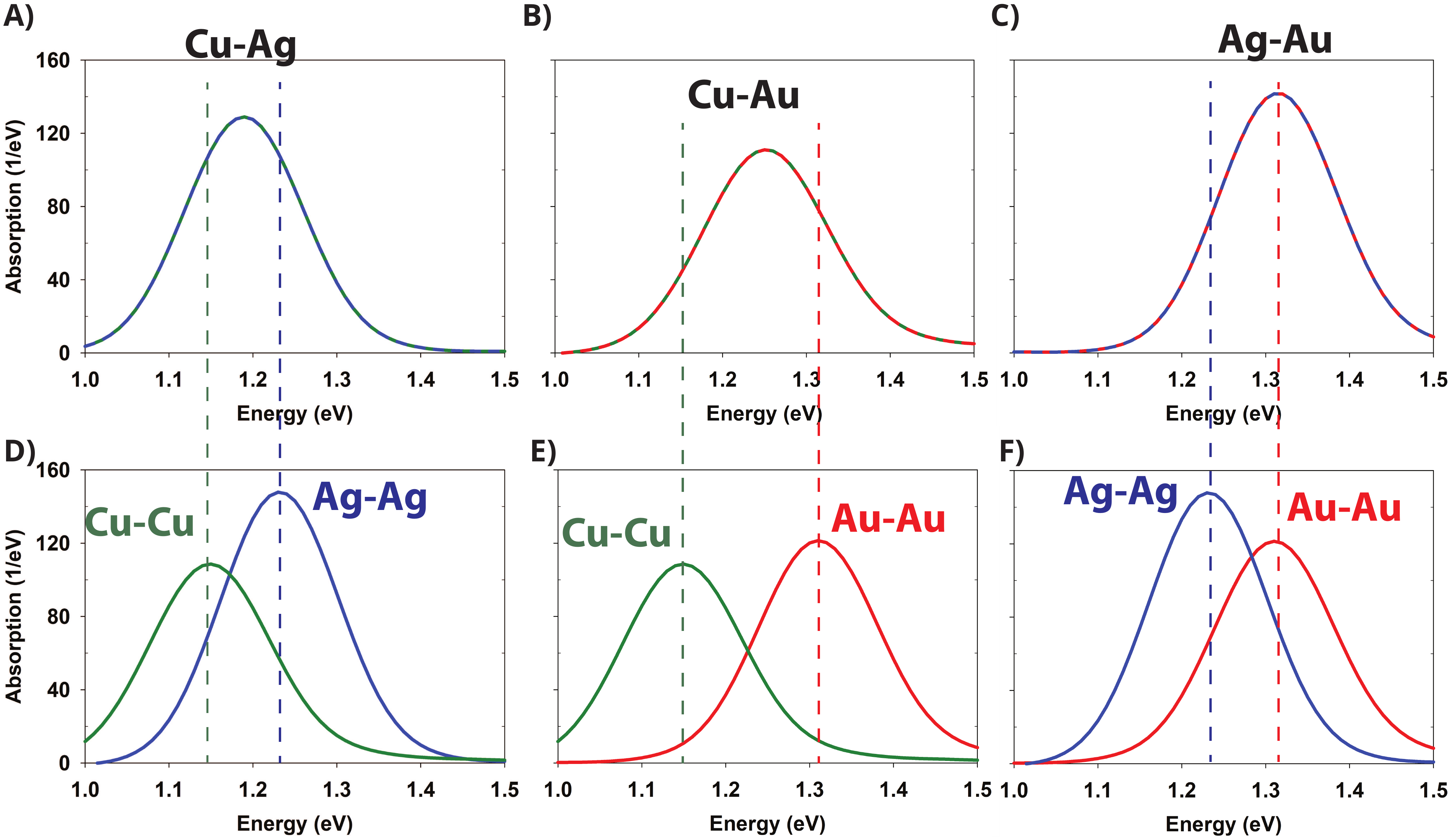}
\caption{Photoabsorption spectra for mixed arrays comprising two chains of 14 atoms of a) Cu-Ag, b) Cu-Au, and c) Ag-Au double chains. d-f) Spectra for the pure Cu-Cu, Ag-Ag, and Au-Au 2~x~14 double chains. }
\label{fig:TMrow_array_spectra}
\end{figure}

\begin{figure}
\includegraphics[width=0.5\textwidth]{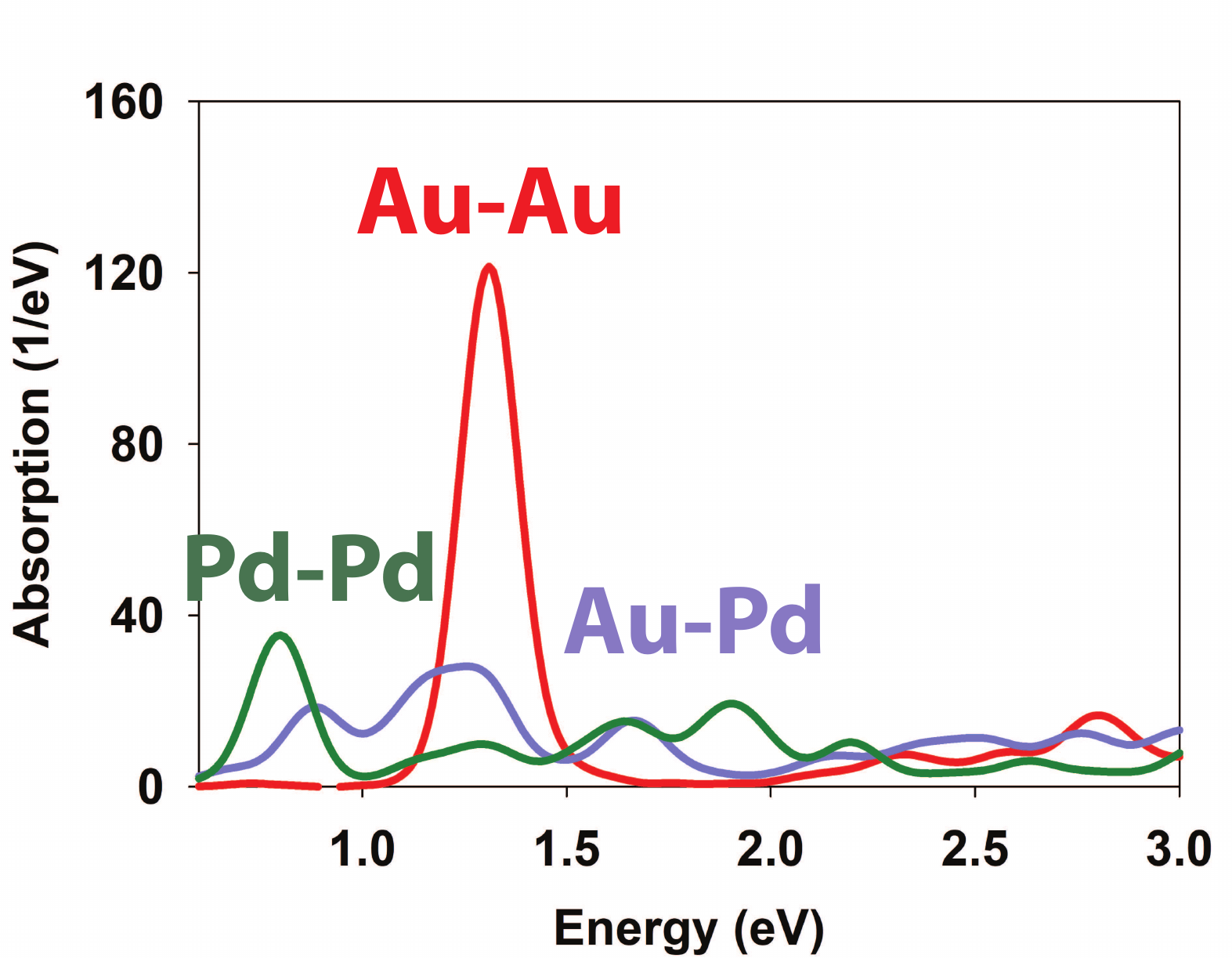}
\caption{Photoabsorption spectra of the mixed double chain array comprising one chain of 14 Pd and one chain of 14 Au atoms, and pure Pd-Pd and Au-Au 2~x~14 double chains. }
\label{fig:PdAu_array_spectra}
\end{figure}

In this sub-section, we focus on mixed, or coupled, arrays comprising two chains each fourteen atoms long. By mixing chains of different elements, the position of the surface plasmon resonance can be tuned or the molecular plasmon fragmented into several peaks with broadband photoabsorption. For example, as seen in Figures~\ref{fig:TMrow_array_spectra}(a,d) the plasmon peak of 14Ag-14Cu double chain is situated halfway between the plasmon peak of pure Ag and pure Cu double chains. However, when the Cu chain is joined with the Au chain, the plasmon peak is shifted toward that of the pure Au double chain (Figures~\ref{fig:TMrow_array_spectra}(b,e)), and when the Ag chain is joined with the Au chain, the plasmon peak is practically at that of the pure Au double chain (Figures~\ref{fig:TMrow_array_spectra}(c,f)). In striking contrast, when a Au chain is joined with a TM (Pd, Pt, Ni, Fe, Rh) chain, the plasmon modes are highly suppressed and several peaks form as seen in the case of the 14Au-14Pd double chain in Figure~\ref{fig:PdAu_array_spectra}. 

The behavior of the molecular plasmon in the double chains comprised of two different noble metals, which are shown in Figure~\ref{fig:TMrow_array_spectra}, are first explained by the electron confinement depending on the atomic ionization potentials and atomic size. Thereafter the trends are further analyzed using TCMs. The result for the 14Au-14Pd double chain more prominently reflect the chemical properties of the two species and require a quantum mechanical analysis at the outset.  

The effect of mixing arrays with homonuclear chains on the average electron density can be explained using the differences in the decay length of the valence electron density and the ionization potentials. The valence electron density decay length of Cu is smaller than Ag and Au, which are approximately equal.~\cite{clementi1967atomic} The ionization potentials of Au is much larger than Cu and Ag, which are approximately the same. Combining these properties, we can expect that the valence electron density between the chains in the Cu-Ag mixed array is approximately equal to the average of the valence electron densities of the pure Cu double chain array and the pure Ag double chain array. Therefore the plasma frequency for the mixed Cu-Ag double chain, which is proportional to the square root of the average electron density, is located at the midpoint between the plasmon of the Cu-Cu and Ag-Ag double chains. In the case of the Cu-Au double chain, the larger ionization potential of Au reduces the electron spillout from the Cu chain in comparison with the Cu-Cu double chain. However, since the decay length of the electron density of the Cu chain is smaller and the interchain distance is fixed, the resulting electron density between the Cu and Au chains does not rise very much from the average between between two Cu and two Au chains. Therefore the plasma frequency is shifted modestly from the midpoint between the Cu and Au double chain values. Finally, in the case of the Ag-Au double chain the decay length of the electron densities are similar for both chains. Then the larger ionization potential for Au can compensate the smaller valence electron contribution from the Ag chain by reducing the electron spillout from the Ag chain and as a result the plasma frequency for the Ag-Au double chain is nearly the same as for the Au-Au double chain.

\begin{figure}
\includegraphics[width=0.32\textwidth]{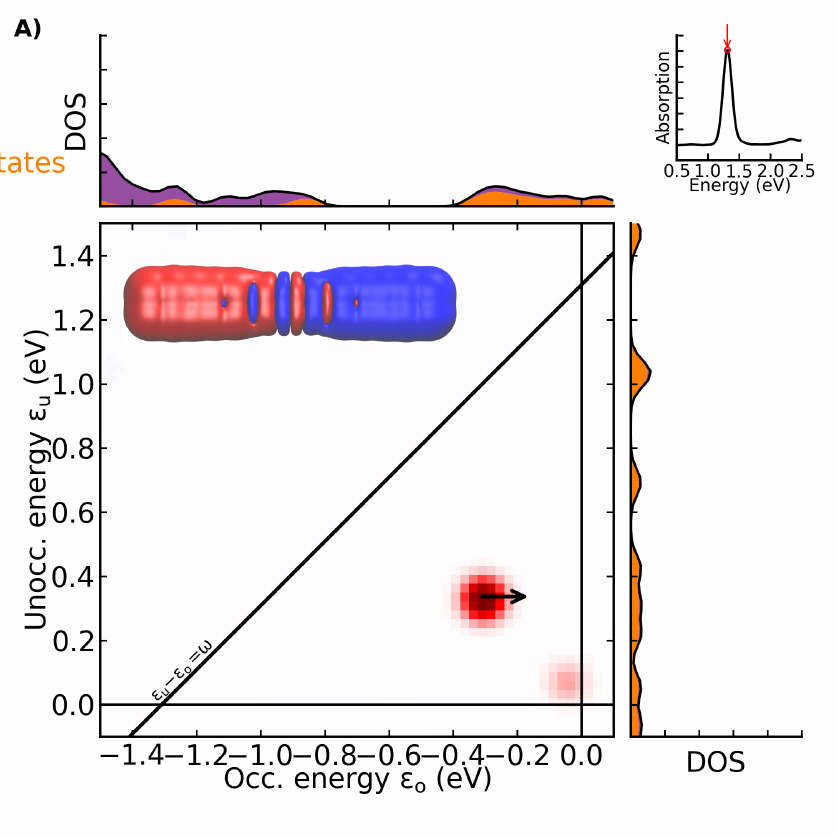}
\includegraphics[width=0.32\textwidth]{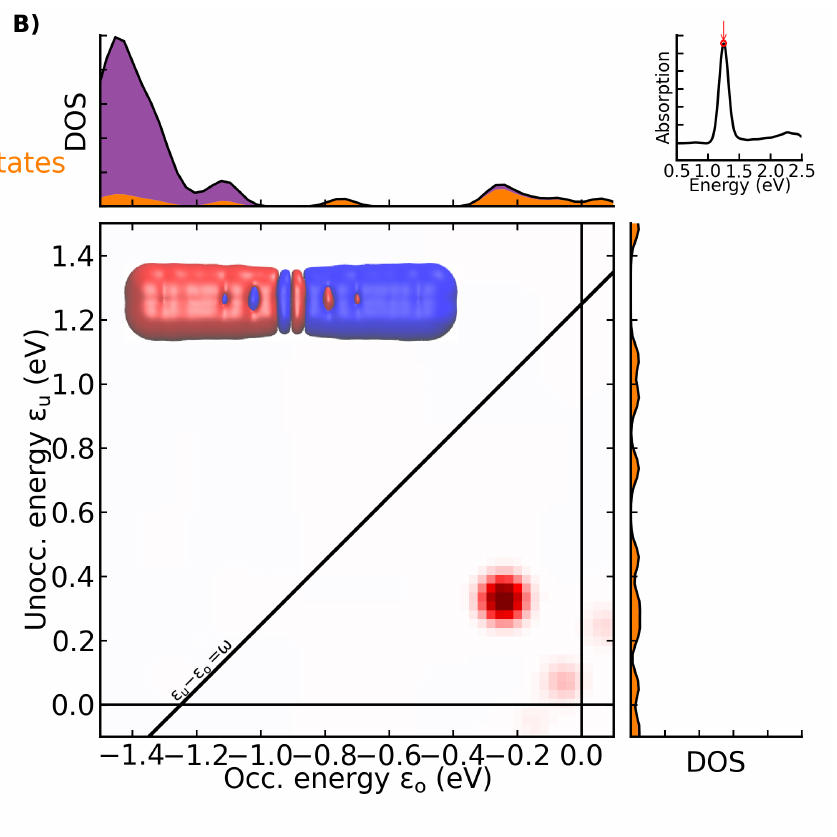}
\includegraphics[width=0.32\textwidth]{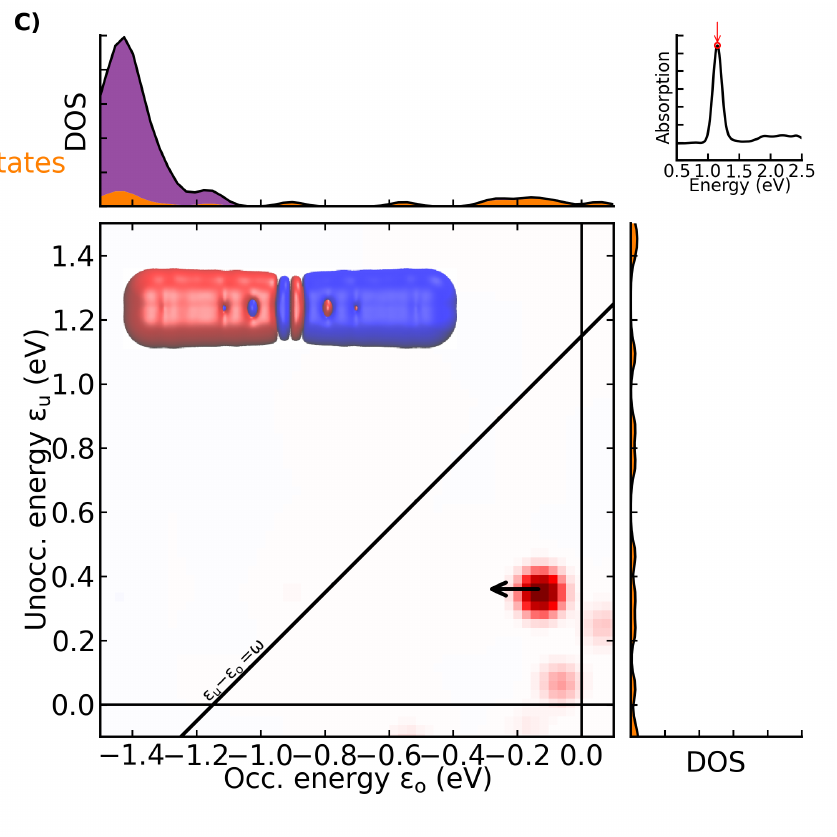}
\caption{Transition contribution maps for plasmonic excitations in a) Au-Au, b) Au-Cu, and c) Cu-Cu 2~x~14 double chains at $\omega$ = 1.31, 1.25, and 1.15 eV, respectively.}
\label{fig:array_TCM_cumixed}
\end{figure}

We will now turn to the more detailed description based on TCMs which are shown in Figure~\ref{fig:array_TCM_cumixed} for Cu-Cu, Cu-Au, and Au-Au double chain arrays. In each case, the plasmon excitation has one dominant KS transition between delocalized sp-states (see the KS orbitals in Figure~\ref{fig:2xpureTCM}a) and the strong dipolar induced density is similar to the pure Au double chains. In this picture series, the shift of the plasmon energy is related to the energy of its constituent dominating KS transition. For example in a mixed Cu-Au array, seen in Figure~\ref{fig:array_TCM_cumixed} middle, the occupied state of the collective excitation is shifted midway between that of a pure Au and and pure Cu array. The KS orbitals of the array can be thought of as a construction of the KS orbitals of two individual chains which form a bonding and an anti-bonding orbital. In the mixed array, the energy mismatch between the KS orbitals of Au and Cu chains produces a bonding KS orbital energetically between the KS orbitals of the pure Cu and pure Au arrays. The unoccupied states, formed by a bonding KS orbital with an additional node, remain at approximately the same energy. Thus the plasmon resonance energy of the mixed array is in between those of the corresponding pure arrays. The TCMs of Ag-Au and Cu-Ag mixed double chains are similar to the Au-Cu mixed arrays which are shown in Figures S4 and S5.

\begin{figure}
\includegraphics[width=0.45\textwidth]{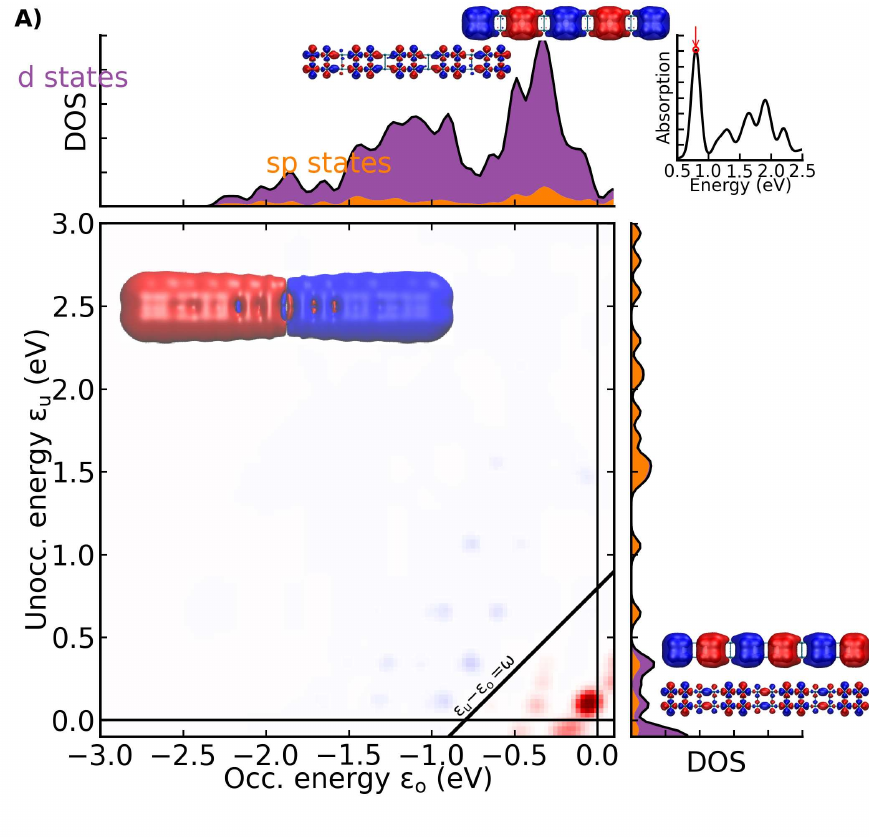}\\
\includegraphics[width=0.3\textwidth]{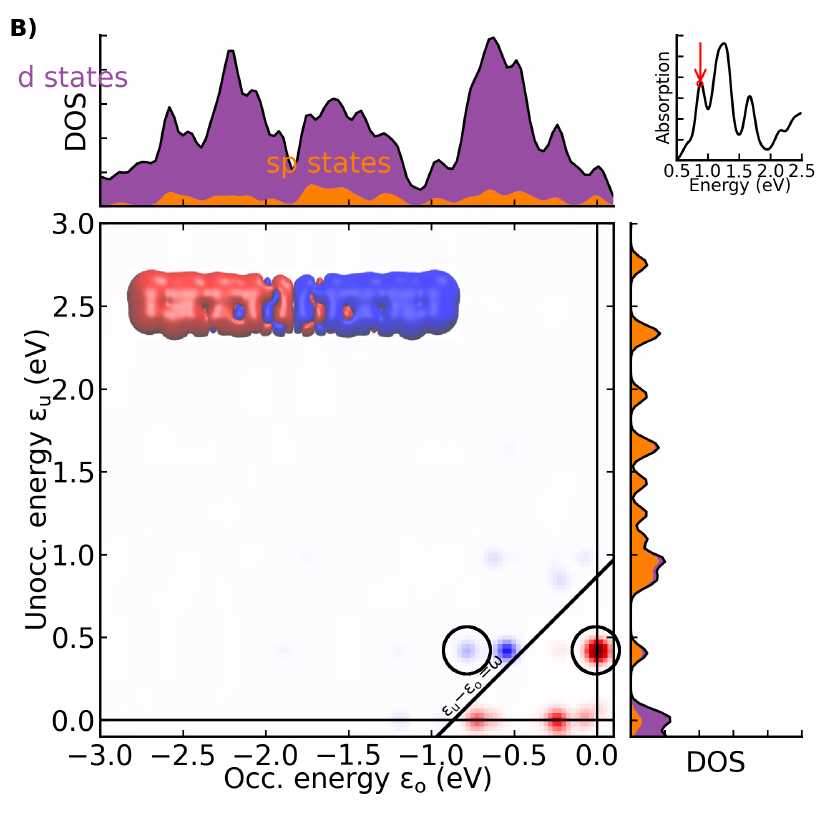}
\includegraphics[width=0.32\textwidth]{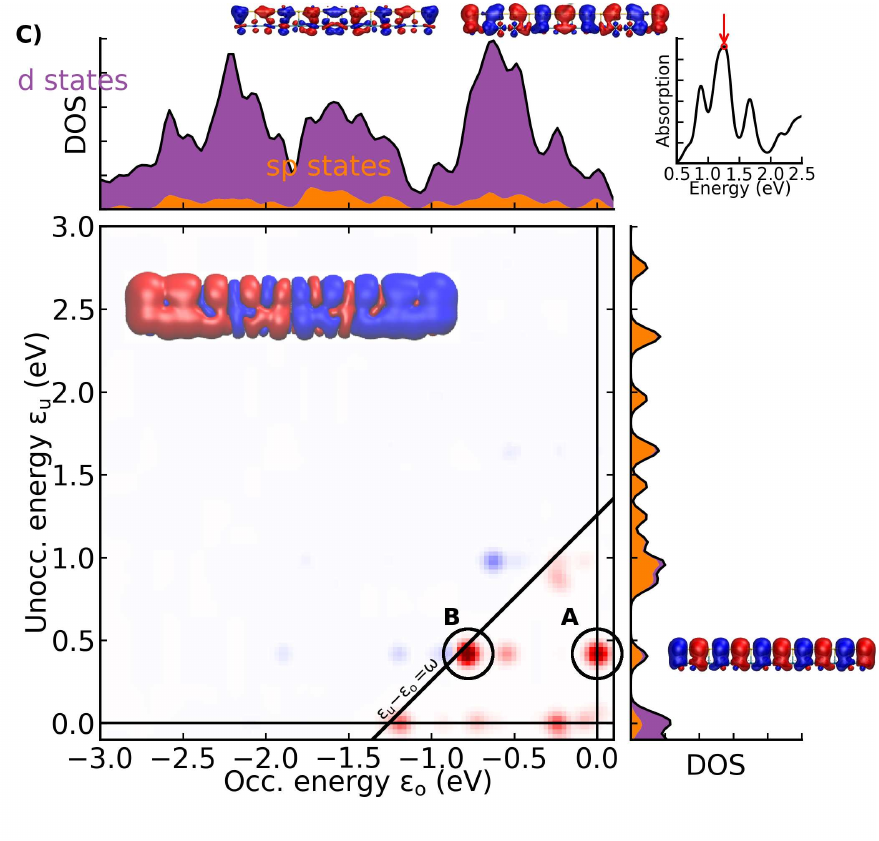}
\includegraphics[width=0.3\textwidth]{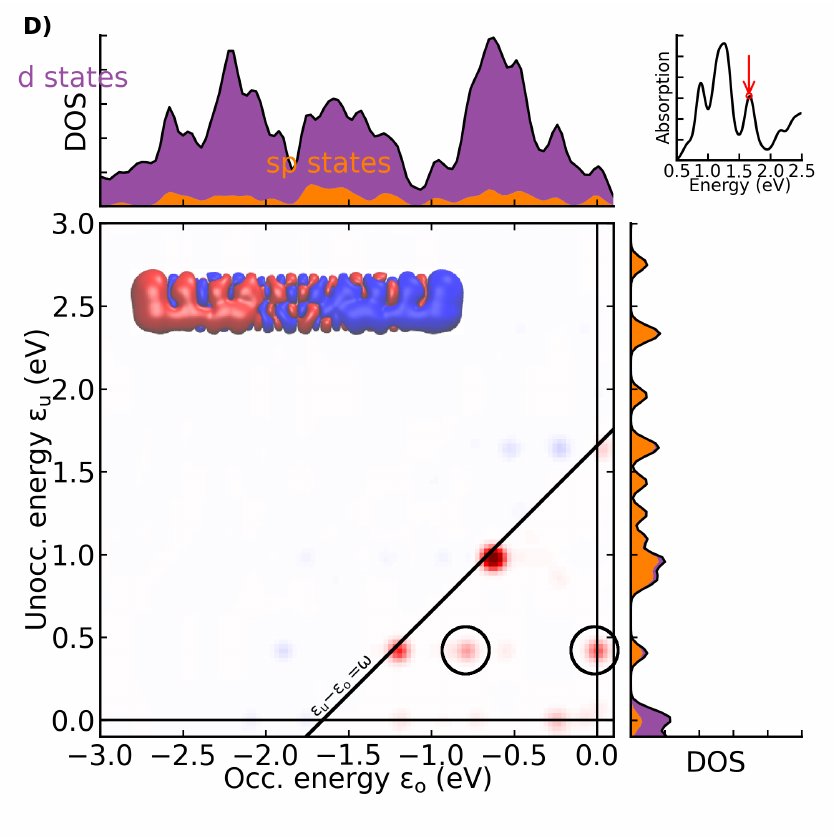}\\
\includegraphics[width=0.35\textwidth]{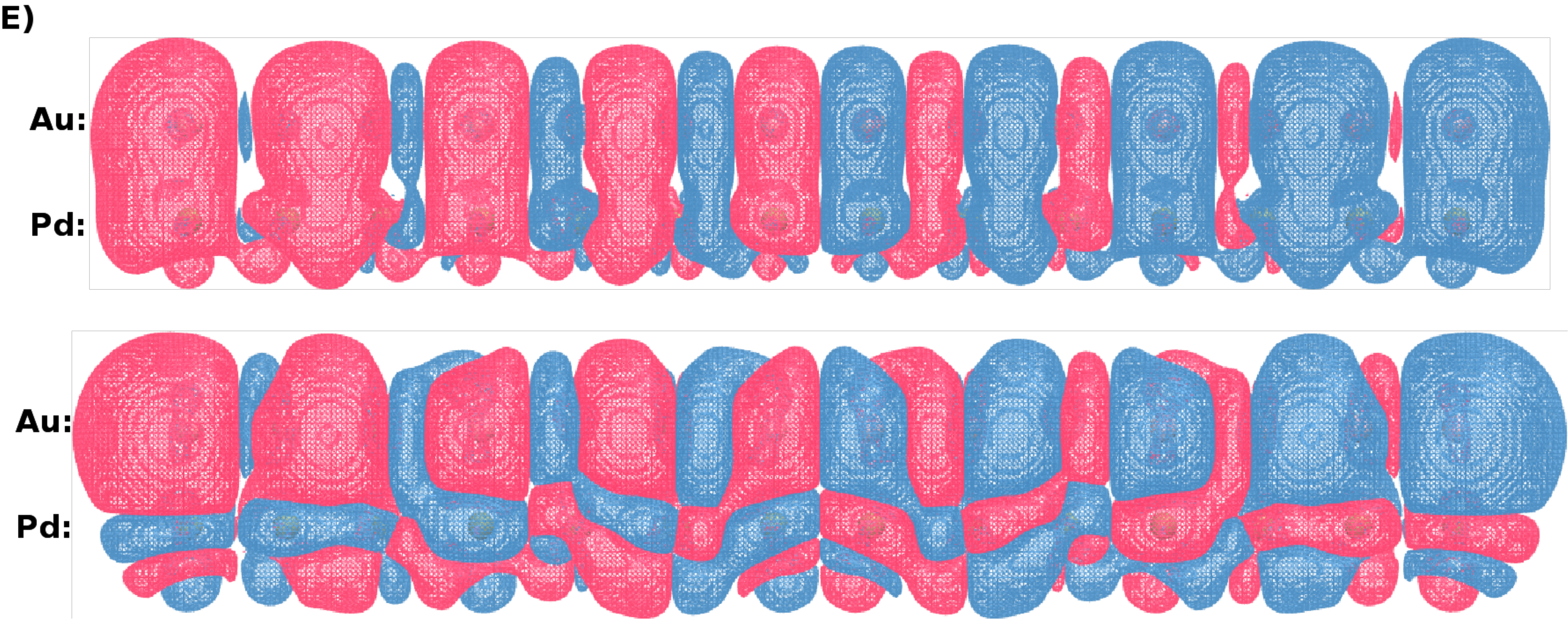}\\
\caption{\textbf{Plasmons from pure Pd-Pd and mixed Pd-Au arrays.} a) Contribution to the plasmonic excitation at $\omega$ = 0.80 eV of a Pd-Pd array. b-d) - Contribution to the absorption peak of a Pd-Au mixed array at $\omega$ = 0.87, 1.26, and 1.66 eV, respectively. e) - Isolated transition densities of the transitions A (top) and B (bottom) which are formed from the KS orbitals shown next to the DOS in (c).}
\label{fig:TCM_TMrow}
\end{figure}

Figure~\ref{fig:PdAu_array_spectra} shows a rather intense absorption peak at 0.80 eV for the 2~x~14 Pd double chain. The TCM for this excitation in Figure~\ref{fig:TCM_TMrow} shows a structure with plasmonic characteristics, i.e., all KS transitions contributing strongly to the excitation are well below the $\omega$ probe line. Especially, the transition just inside the corner formed by Fermi levels and marked by the red spot contributes to the collective excitation. In spite of the high energy d levels, the array is able to sustain such a plasmonic excitation, but this may not be the case in other array sizes. 

The absorption spectrum of the Pd-Au double chain in Figure~\ref{fig:PdAu_array_spectra} is significantly suppressed and fragmented compared to the pure and mixed noble metal double chains. The fragmentation is due to both sd-hybridization and transient d-transitions between the overwhelming amount of occupied d-states near the Fermi level introduced by Pd. The d-states can be seen in the DOS of the Au-Pd mixed array in Figures~\ref{fig:TCM_TMrow}(b-d). While the fleeting transitions between d-states contribute to the photabsorption only near the probe line, some of the introduced d-states continue to contribute at higher energies. These d-states have hybridized with the sp-delocalized states to form bonding and anti-bonding KS orbitals. 

TCMs conveniently show that the contribution of these individual transitions between sd-hybridized states changes sign near the $\omega$-probe line. It is illustrative to identify which KS transitions cause the new resonances. For example, the two largest contributions to the photoabsorption peak at $\omega$ = 1.26 eV occur from two KS transitions labelled A and B in Figure~\ref{fig:TCM_TMrow}c. Transitions A and B have the same final state, which has 9 nodes, but different initial states. Both initially occupied states have 8 nodes on the Au chain, but they have hybridized with the either the delocalized d$_{yz}$ (in A) or d$_z^2$ (in B) states on the Pd chain. The different symmetry on the Pd chain produces partial induced densities where the Au chain is either in phase with the Pd chain (transition A) or out of phase (transition B) as seen in Figure~\ref{fig:TCM_TMrow}e. Due to the longitudinal nodal difference of one, both partial induced densities have an overall dipole. There is pronounced screening from the localized d-states in both partial induced densities. 

Each KS transition contributes positively or negatively, and a change in sign corresponds to a new photoabsorption resonance. The absorption peak at $\omega$ = 0.87 eV is produced by a change in contribution of transition B and the adjacent d-transition from positive to negative as the frequency decreases (that is, moving from Figure~\ref{fig:TCM_TMrow}c to b). Similar to a linear combination, changing the sign of the individual \textit{ia}-transitions constructs new absorption resonances with unique induced densities. We note that other resonances, such as at $\omega$ = 1.16 or 1.66 eV as seen in Figure~\ref{fig:TCM_TMrow}d, are produced by changes of the contribution in other active KS transitions. The numerous closely spaced transitions effectively broadens the absorption band. In this manner, the sd-hybridization fragments the plasmon by activating additional KS transitions leading to the additional peaks in the photoabsorption spectra.~\cite{Yannouleas1989, rossi2017kohn} As the resonance energy reduces, the total induced density becomes more dipolar as a result of the constructive interference of the dipole moments (see Figure~\ref{fig:TCM_TMrow}(b-d) insets). 

Similar effects are observed in other Au-TM mixed arrays with d levels near the Fermi level. The spectra and TCMs of pure Pt double chain arrays and 14Au-14Pt mixed arrays are shown in Figures S6 and S7, respectively.

\section{Conclusions} \label{conc}
We have analyzed the optical properties of multiple pure and mixed noble/TM metal planar nanoarrays of fixed length and varying width. The absorption peak moves to higher (visible light) energies when the width of the array increases, but the developing subband structure, due to the quantum confinement in the finite atomic arrays, produces a non-monotonic shift. The electron spillout is strongest in the smaller arrays, and clearly expressed in a redshift in the absorption mode between arrays with three and four Au chains, and their respective work functions.

The strong absorption mode is produced by low-energy transitions between sp-delocalized states with large dipole moment, and TCMs provide insight into their plasmonic nature. The contributions to the absorption mode fall below the $\omega$ probe line, which is present even in a single Au chain, indicating molecular-plasmonic character of the mode. As the width of the nanoarray is increased, the number of transitions originating from the newly opened subbands becomes more numerous. The increased number of transition contributions and developing nodal structure in atomic arrays is instructive in understanding the behavior of larger nanoparticles. Although nodal structure is more complex in larger nanoparticles, it is evident that the plasmons of larger nanoparticles have contributions from an increasing amount of KS transitions and some of which may fragment the plasmon~\cite{Yannouleas1989, rossi2017kohn}. 

The energy position of the d-electron levels affects the optical properties of mixed arrays of homonuclear chains. The intense plasmon is either maintained when mixing noble metal chains, or fragmented when mixing a noble metal chain with a TM chain. The sd-hybridization and introduced d-states activate additional KS transitions to produce new resonances. These resonances fragment and diminish the photoabsorption intensity. We have mainly focused on Au chains mixed with Pd or Pt atoms, but as pure Ag and Cu systems behave similarly to pure Au system, one can expect similar picture when Ag and Cu are doped with TM atoms. Provided that such multi-chain systems are built experimentally on a ``neutral'' substrate, the results and understanding obtained in this paper may be applicable to different optical technologies.

\begin{acknowledgement}
The work was supported in part by DOE Grant DE-FG02-07ER46354 (TR) and as part of the Academy of Finland Centre of Excellence program (project 312298) (MP).
\end{acknowledgement}

\bibliography{dopant_bib}

\end{document}